\documentclass[12pt]{article}
\usepackage{epsfig}

\usepackage{amssymb}
\usepackage{amsmath}
\usepackage{amsfonts}

\def\be{\begin{equation}}
\def\ee{\end{equation}}
\def\ba{\begin{array}{c}}
\def\ea{\end{array}}

\def\ben{$$}
\def\een{$$}

\newcommand{\bea}{\begin{eqnarray}}
\newcommand{\eea}{\end{eqnarray}}

\newcommand{\bbr}{\br\!\br}
\newcommand{\kkt}{\kt\!\kt}

\newcommand{\kt}{\rangle}
\newcommand{\br}{\langle}
\begin{document}

\titlepage

\vspace{.35cm}

 \begin{center}{\Large \bf

Fundamental length in quantum theories with ${\cal PT}-$symmetric
Hamiltonians

  }\end{center}

\vspace{10mm}

 \begin{center}

 {\bf Miloslav Znojil}

 \vspace{3mm}
Nuclear Physics Institute ASCR,

250 68 \v{R}e\v{z}, Czech Republic

{e-mail: znojil@ujf.cas.cz}

\vspace{3mm}


\end{center}

\vspace{5mm}


\section*{Abstract}

One-dimensional motion of a quantum point particle is usually
described by its wave function $\psi(x)$ where the argument $x\in
I\!\!R$ represents a (measurable) coordinate and where the
integrated probability density is normalized to one, $\int
\psi^*(x)\psi(x)=1$. The direct observability of $x$ may be lost in
${\cal PT}-$symmetric quantum mechanics where a ``smeared" metric
kernel $\Theta_{(x,x')}\neq \delta(x-x')$ may enter the
double-integral normalization, $\int \int
\psi^*(x)\Theta_{(x,x')}\psi(x')=1$. We argue that such a formalism
proves particularly suitable for the introduction of a nonvanishing
fundamental length $\theta>0$ which would characterize the
``smearing width" of the kernel $\Theta_{(x,x')}$. The technical
feasibility of such a project is illustrated via a toy family of
Hamiltonians $H^{(N)}(\lambda)$ taken from paper I (M. Znojil, Phys.
Rev. D 78 (2008) 025026). For each element of this family the
complete set of all the eligible metric kernels
$\Theta^{(N)}_{(x,x')}(\lambda)$ is constructed in closed form. We
show that at any pre-selected non-negative fundamental length these
metrics can be made to vanish unless $|x-x'|\leq \theta$. The
strictly local inner product of paper I recurs at $\theta=0$ while
the popular ${\cal CPT}-$symmetric option requires $\theta=\infty$
in this language.

\newpage
 \section{Introduction \label{zacatek} }

The introduction of a minimal length scale (denoted, say, by
symbol $\theta$) is tempting on empirical as well as purely
pragmatic grounds. Typically, its existence would facilitate the
regulation of the high energy asymptotics in field theory
\cite{prva15}, etc. A more ambitious motivation of its
introduction might be sought in M theory or string theory in their
various limiting cases \cite{paty12}. Connections between
$\theta>0$ and the emergence of certain singularities with
nontrivial physical meaning could further be sought in
cosmological applications of quantum theory \cite{druha9}. In the
astrophysical context, last but not least, the fundamental length
might be identified as accounting for the dark energy \cite{dvoj2}
or for the inflationary era in the early evolution of the universe
\cite{dalsi1}.

In the majority of similar considerations the fundamental length
scale emerges as a free parameter. For the particular quantum
dynamics of spacetime, for example, its value can be related not
only to the Planck length but also, say, to a nonvanishing
cosmological constant or to the vacuum energy density
\cite{ctvrty6}. In the simplified context of quantum mechanics this
quantity can even be treated as one of phenomenological,
experimentally determined characteristics, say, of a
condensed-matter system \cite{last3}. In parallel, in some more
ambitious theoretical studies the introduction of a fundamental
length constant is being based on a deeper principle like the
stabilization requirement imposed upon relativistic algebra
\cite{treti5}.

Many of the latter ideas are implemented using the assumption of
non-commuta-tivity of coordinates (cf. a rather nonsystematic
selection \cite{treti5,Frikkie} of some sample references). In what
follows we intend to develop a different theoretical concept in
which the existence of fundamental length will find its origin and
realization via analytic rather than algebraic considerations. We
shall particularly be guided by the recent innovative
interpretations of certain analytic potentials characterized by
their so called ${\cal PT}-$symmetry (for a review, paper
\cite{Carl} is recommended).

More specifically, in our preparatory section \ref{sII} we shall
define ${\cal P}$ and ${\cal T}$ via a sample Hamiltonian $H$
(paragraph \ref{buslie}) and explain why we believe that in quantum
theory  of similar models one of the most natural definitions of the
length-scale $\theta>0$ could be based on a suitable particular
realization of the physical Hilbert space ${\cal H}$. We show, in
paragraph \ref{busliebe}, how a ``smearing of coordinates" emerges
as allowed by the well known ambiguity of the physical inner
products in ${\cal H}$. This enables us to conjecture that in
general, one should be able to remove or at least suppress this
ambiguity by the requirement that the range of the smearing of
coordinates acquires precisely the  pre-selected non-negative value
$\theta$.

For a quantitative understanding of the similar quantum models with
built-in scale $\theta$ a family of very specific illustrative
examples is introduced in section \ref{toynbee}. The feasibility of
their analysis is achieved not only by the use of a non-perturbative
technique of solving Schr\"{o}dinger equations (based on a
discretization of coordinates, cf. paragraph \ref{sec3a}) but also
by the choice of a very elementary, next-to-trivial interaction (cf.
ref. \cite{prd} or eq.~(\ref{RuKaBaC}) in paragraph \ref{prevodnik}
below). In this way all our Hilbert spaces become finite-dimensional
and the Hamiltonians become represented by certain $2K-$dimensional
matrices $H^{(2K)}(\lambda)$ where the real parameter $\lambda$
controls their non-Hermiticity. At the minimal Hamiltonian-matrix
dimension $2K=N=2$ this renders {\em all} the eligible ``physical"
inner products in ${\cal H}$ available in closed form  (paragraph
\ref{Nje2}) making the discussion of the fundamental length $\theta$
trivial.

The $N=2$ conclusions encouraged us to develop and employ a
linear-algebraic algorithm of the construction of metrics applicable
at any even Hilbert-space dimension $N=2K$. The
symbolic-manipulation results are sampled, in section
\ref{mquattro}, at the two subsequent integers $K = 2$ and~$3$. The
availability as well as unexpectedly transparent matrix structure of
the resulting matrices of the metric proved crucial for our present
fundamental-scaling purposes. In essence we revealed that the set of
$\Theta^{(N)}$ appears composed of subsets in which the matrices
$\Theta^{(N)}$ acquire a band-matrix structure. The elementary
length~$\theta$ (i.e., the range of the smearing of coordinates in
the inner product) is then identified with a measure of
non-diagonality (i.e., with the ratio between the number of
diagonals and dimension) of these matrices.

In section~\ref{IVaauv} we turn our attention, first of all, to the
next few higher dimensions $N \geq 8$ at which the computer-assisted
brute-force determination of  {\em all} of the eligible ``physical"
inner products in ${\cal H}$ still remains feasible. Of course, the
very enumeration of the $N-$parametric sets of the resulting
matrices $\Theta^{(N)}$ becomes clumsy. For this reason we developed
a recurrent technique of their description which offers their
compact classification as well as their explicit and compact
description at any dimension $N=2K$.

The phenomenological core of our present  message lies in several
observations formulated in several separate subsections of section
\ref{uvod}. We point out, i.a., that our present choice of the
family of Hamiltonians $H^{(2K)}(\lambda)$ looks particularly
suitable for illustrative purposes since the bound-state energies
remain real in a dimension-independent interval $\lambda\in (-1,1)$.
One finds out that at a fixed dimension $N=2K$  the
phenomenologically motivated fundamental length $\theta$ can equally
well be perceived as a means of  classification of eligible metrics.
This feature of our solvable model becomes particularly useful when
a {\em fixed choice} of the fundamental length $\theta>0$ is
analyzed in the continuous-coordinate limit $N \to \infty$.

Our last, summary section \ref{summary} re-emphasizes the importance
of the exact, non-perturbative solvability of our schematic
benchmark example which admits the exhaustive and exact construction
of the {\em complete menu} of matrices $\Theta^{(N)}$ of the metrics
at a given $N$. In the future, having such a transparent methodical
guide at our disposal we may expect that all the prospective
transitions to the more realistic interaction models will be
facilitated and/or more easily mediated via approximate (e.g.,
perturbation-theory) techniques.

\section{Nonlocal inner products \label{sII}}

\subsection{${\cal PT}-$symmetric models \label{buslie}}

Our main source of inspiration can be traced back to the discoveries
of existence of the {\em real} spectra of energies generated by
non-Hermitian one-dimensional Hamiltonians \cite{Caliceti,BG}. For
illustration let us recollect just the Buslaev's and Grecchi's (BG,
\cite{BG}) Hamiltonian $H= p^2+V(x)\neq H^\dagger$ of this type
containing the asymptotically quartic ``wrong-sign" potential
 \ben
 V(x)=V^{(BG)}(x) = -(x-{\rm i}\varepsilon)^4 + {\cal O}(x^2)
 \neq V^*(x)
 \,,
 \ \ \ \ \ x \in (-\infty,\infty)\,,
 \ \ \ \ \ \varepsilon>0\,.
 \een
The distinguishing feature of this non-Hermitian Hamiltonian
(studied, later, also by Jones et al \cite{Mateo}) is that it is
characterized by its ${\cal PT}-$symmetry
 $
 {\cal PT}H = H{\cal PT}
 $ where the operators ${\cal P}$ and  ${\cal T}$
stand for the parity and time reversal, respectively.

The essential merit of the BG model is that the asymptotically
dominant parts of the general solutions $\psi^{(BG)}_{1,2}(x)$ of
the related differential Schr\"{o}dinger equation near the
respective endpoint coordinates $x_{1,2}=\pm \infty$ are easily
deduced,
 \ben
 \psi^{(BG)}_{1,2}(x) = c_+^{(1,2)}\psi_+^{(BG)}(x)
 + c_-^{(1,2)}\psi_-^{(BG)}(x)\,,
 \ \ \ \
 \psi_\pm^{(BG)}(x)
  = e^{\pm \frac{1}{3} {\rm i} x^3 \pm \varepsilon x^2
 +{\cal O} (x)}\,.
 \een
As long as $\varepsilon>0$ we can set $c_+^{(1,2)}=0$ guaranteeing
that the resulting functions $\psi^{(BG)}_{1,2}(x)$ will
asymptotically vanish, $\psi^{(BG)}_{1}(+\infty)=0$ and
$\psi^{(BG)}_{2}(-\infty)=0$. After analytic continuation their
matching near the origin gives the physical bound-state solution
which is analytic at all $x \in (-\infty,\infty)$ and quadratically
integrable, i.e., $\psi^{(BG)}_n(x) \in I\!\!L^2(I\!\!R)$ or
 \be
 \int_{I\!\!R} dx\,
 \left [\psi^{(BG)}_n(x)
 \right ]^*\,\psi^{(BG)}_n(x)\ \ < \ \ \infty\,.
 \label{jednac}
 \ee
It has also rigorously been proved in \cite{BG} that the energies
$E=E_n$ are all real, non-degenerate and growing with the main
quantum number $n = 0, 1, \ldots$.

From our present point of view, it is more important that the
related wave functions remain mutually non-orthogonal. This means
that the physical information carried by these wave functions
remains unclear. In order to restore the physical probabilistic
interpretation of such a ${\cal PT}-$symmetric model one must modify
the inner product \cite{Carl,Ali}. Usually, this goal is achieved by
the replacement of  the unphysical Hilbert space $I\!\!L^2(I\!\!R)
:={\cal H}^{(F)}$ (where F stands for ``first" or ``friendly" or
``false") by its amendment ${\cal H}^{(S)}\neq I\!\!L^2(I\!\!R)$
where S means ``second" or ``standard" and where the bound states
become mutually orthogonal \cite{SIGMA}.

\subsection{Fundamental length $\theta$ as
a measure of non-locality\label{busliebe}}

During the return to the ``standard" Hilbert space ${\cal H}^{(S)}$
one reveals that our illustrative potential $V^{(BG)}(x)$ is an
extremely ``user-friendly" interaction. In this sense many of its
properties appear rather exceptional (cf. \cite{BG} where one finds
that the transition to ${\cal H}^{(S)}$ leaves the interaction
local). For this reason it makes sense to recall also several other
illustrative models. In order to make the picture comparatively
complete one must recollect, e.g., the imaginary cubic oscillator of
Bessis and Zinn-Justin \cite{DB} and/or the whole one-parametric
family of its generalizations with $V(x)/x^2= ({\rm i}x)^\delta$
where $\delta\geq 0$ (cf. \cite{Carl} for more details). In all of
these cases the transition to ${\cal H}^{(S)}$ makes the interaction
strongly nonlocal (for illustration we recommend to check the
details, e.g., via their perturbative illustration at $\delta=1$ in
paper \cite{cubic}).

Let us emphasize that the necessary condition of the possibility of
the transition from the unphysical Hilbert space ${\cal H}^{(F)}$ to
its physical parallel ${\cal H}^{(S)}$ lies in the reality of the
spectrum of the ``non-Hermitian" Hamiltonian in question. Once this
condition is satisfied, the most common realization of the
correspondence ${\cal H}^{(F)} \to {\cal H}^{(S)}$ usually (cf.
\cite{Carl}) proceeds under the assumption that the space ${\cal
H}^{(S)}$ remains spanned by the same wave functions and that only
the inner products are re-defined as non-local \cite{Geyer}. Here we
shall follow the same recipe. Without getting too deeply in the
underlying mathematics let us only recollect that for all the
sufficiently elementary Hamiltonians the inner product used in
${\cal H}^{(S)}$ may be understood as leading to the generalized,
double-integral orthonormalization rule
 \be
 \int_{I\!\!R^2}  dx\,dx'\,\psi_m^*(x)\,\Theta_{(x,x')}\,\psi_n(x')
 =\delta_{mn}\,
 \label{laots}
 \ee
where $\delta_{mn}$ is Kronecker symbol. For the Hamiltonians which
are ``trivially" Hermitian in ${\cal H}^{(F)}$ the standard textbook
scenario characterized by the Dirac's metric
$\Theta_{(x,x')}=\delta(x-x')$ leads merely to the degenerate
version eq.~(\ref{jednac}) of eq.~(\ref{laots}). {\it Vice versa},
Dirac-non-Hermitian Hamiltonians $H \neq H^\dagger$ with real
spectra necessitate a selection of a Hamiltonian-dependent metric
kernel $\Theta_{(x,x')}\neq \delta(x-x')$ in eq.~(\ref{laots}). One
just replaces the elementary Dirac's Hermitian conjugation
 \ben
 {\cal T}^{(F)}: \psi(x) \to \psi^*(x)
 \een
by its non-local, more complicated version
 \be
  {\cal T}^{(S)}: \psi(x) \to
  \int_{I\!\!R}  dz\,\psi^*(z)\,\Theta_{(z,x)}\,.
   \label{polaotse}
 \ee
There exist examples (based on a sufficiently elementary choice of
Hamiltonian $H$ - cf., e.g., \cite{David}) where the metric
$\Theta_{(x,x')}$ itself can even be constructed exactly,
non-perturbatively .

In this language our present key message is that our
equations~(\ref{laots}) and (\ref{polaotse}) may be complemented by
the phenomenologically motivated limited-range constraint
 \be
 \Theta_{(x,x')} \neq 0 \ \ {\rm only \ if}\ \ |x-x'| \leq \theta\,
 \label{laotse}
 \ee
using any preselected real quantity $\theta>0$. One must keep in
mind that this quantity is just an elementary upper estimate of
non-locality imposed upon the metric in ${\cal H}^{(S)}$. Its size
may be perceived as a measure of the ``smearing" of the coordinate
$x$ which is due to the loss of the direct physical meaning and
measurability of the real variable $x$ in ${\cal H}^{(F)}$.

Strictly speaking one should not even call quantity $\theta>0$ a
``length". At the same time, one feels that once the range of the
smearing of the metric kernel is assumed restricted by
eq.~(\ref{laotse}), the effect of this smearing will quickly
decrease when the measured distances exceed the preselected
``fundamental" value $\theta$. This observation may be also read as
a core of our present project specifying a class of models where the
smearing is guaranteed to be safely short-ranged in $\theta-$scaled
units.

In this setting the mathematical questions emerge which concern not
only the existence of similar models (this question will be answered
here affirmatively and constructively) but also their further
properties. In this sense, our present paper should be perceived as
the first step towards a more extensive theory. Certainly, it will
be non-Hermitian in ${\cal H}^{(F)}$ (where the variable $x$ is only
a non-observable auxiliary quantity) but safely Hermitian in ${\cal
H}^{(S)}$ (of course, only here the fully consistent concept of
distance can be defined). In this context, equation (\ref{laotse})
might acquire its proper meaning as mediator of coexistence between
certain asymptotic locality and short-range non-locality of certain
less standard models of quantum dynamics.

\section{Toy model \label{toynbee} }

The choice of the value of parameter $\theta$ must stay compatible
with our intuition and available experimental evidence. For example,
even the extreme choice of a very large $\theta$ may be tolerated in
the context of bound states where the observable range of
non-locality will effectively be limited by the exponential decrease
of wave functions~\cite{Jones}. In contrast, for a consistent
description of scattering one must necessarily require that the
bound $\theta$ upon the non-locality in eq.~(\ref{laotse}) {\em
must} at least asymptotically be very small~\cite{prd}. In this
sense it is rather encouraging that there exist several different
solvable models of scattering where the smearing size $\theta$ in
eq.~(\ref{laotse}) is strictly equal to zero~\cite{prd,discrete}.
The methodical importance of the latter family of illustrative
examples is further underlined by the fact that for the vast
majority of quantum models with nontrivial metrics the numerical
value of $\theta$ happened to remain infinite \cite{Jones,delta}.

On this background one must be careful with predictions of the
existence of Hamiltonians and metrics for which our upper-estimate
$\theta$ of the built-in non-locality proves nontrivial, i.e.,
non-vanishing and not too large. Thus, we have to offer an explicit,
schematic, constructive example of such a Hamiltonian $H$ and of its
specific short-range-smearing metrics~$\Theta$.

\subsection{Runge-Kutta lattice of coordinates\label{sec3a}}

Let us replace the differential form of a given Hamiltonian
$H=p^2+V(x)$ by its discretized Runge-Kutta approximation leading to
the linear difference Schr\"{o}dinger equation
 \be
  -
 \frac{\psi(x_{k+1})-2\,\psi(x_{k})+\psi(x_{k-1})}{h^2}
  +V(x_k)
 \,\psi(x_{k})
 =E\,\psi(x_k)\,.
 \label{diskr}
 \ee
Using a suitable, finite or infinite cutoff  $L >0$ we set
 \be
 x_{-K}=-L, \
 x_{-K+1}=-L+h, \
 \ldots,
 x_{0}=-\frac{h}{2}, \
 x_{1}=\frac{h}{2}, \
 \ldots,
 x_{K+1}=L \,.
 \label{body}
 \ee
The lattice spacing $h=2L/(2K+1)$ decreases with $N=2K$ in both the
bound-state phenomenological regime (where $L$ should be kept
constant, cf. paper \cite{sqw}) and the scattering regime (where
$L=L(N)$ should grow with $N$, see paper \cite{prd}).

On each level of precision ${\cal O}(h)$ and for virtually  any
strictly local potential $V(x)$ the latter recipe makes the
practical numerical solution of non-Hermitian Schr\"{o}dinger
equations at a fixed lattice-point-distance $h$ decisively
facilitated \cite{thatwork}. The discretization of the coordinates
reduces also the above-mentioned double-integral orthonormalization
rule (\ref{laots}) to its discrete analogue so that the integral
kernels $\Theta_{(x,y)}$ become replaced by matrices
$\Theta_{j,m}^{(N)}(\lambda)$. Under suitable mathematical
assumptions \cite{Geyer} these metric matrices define the inner
product between any two elements $ \psi$ and $\phi$ of our
Runge-Kutta version of the physical Hilbert space of states ${\cal
H}^{(S)}$,
 \be
 \sum_{n=-K+1}^K\,\sum_{n'=-K+1}^K\,
  \psi^*(x_n)\,
 \Theta_{n,n'}^{(2K)}(\lambda)\,\phi(x_{n'}):= \bbr \psi|\phi\kt \,
 \label{newhdi}
 \ee
(cf. \cite{SIGMA} for more details).

\subsection{Minimally non-local interactions of ref. \cite{prd}
\label{prevodnik}}

In our recent studies of scattering \cite{prd,discrete,newest} we
revealed that the finite-range constraint (\ref{laotse}) can be
satisfied (and that one can even easily reach its lower bound
$\theta=0$) provided that a non-locality is admitted in the
potential. A ``minimal" generalization of this type leads to the
following  Runge-Kutta Schr\"{o}dinger equation,
 \ben
  -
 \frac{\psi(x_{k+1})-2\,\psi(x_{k})+\psi(x_{k-1})}{h^2}+
 V(x_k,x_{k+1})\,\psi(x_{k+1}) \ \ \ \ \ \ \ \ \ \ \ \
 \ \ \ \ \ \ \ \ \ \ \ \
 \een
 \be
 \ \ \ \ \ \ \ \ \ \ \ \
 \ \ \ \ \ \ \ \ \ \ \ \
 +V(x_k,x_{k})\,\psi(x_{k})
 +V(x_k,x_{k-1})
 \,\psi(x_{k-1})
 =E\,\psi(x_k)\,.
 \label{diskretni}
 \ee
In principle, eq.~(\ref{diskretni}) must be complemented by
asymptotic  boundary conditions. Keeping in mind, nevertheless, that
the analysis of the scattering scenario has already been performed
in ref.~\cite{prd}, we intend to deal  with the bound-state option
only,
 \be
 \psi(x_{-K})=0\,,\ \ \ \ \
 \psi(x_{K+1})=0\,.
 \label{RKBC}
 \ee
Let us pick up the most elementary nonlocal interaction as
recommended in ref.~\cite{prd} and return to the related
eq.~(\ref{diskretni}) where just the two coupling constants will be
different from zero,
 \be
 V(x_0,x_1)=-\lambda\,,\ \ \ \ \
 V(x_1,x_0)=+\lambda\,.
 \label{RuKaBaC}
 \ee
At each finite $K=1,2,3,\ldots$ or $N=2,4,6,\ldots$ the resulting
Schr\"{o}dinger bound-state eigenvalue problem (\ref{diskretni}) +
(\ref{RKBC})  + (\ref{RuKaBaC}) will degenerate to the
diagonalization of the respective {\em finite} matrix
$H^{(N)}(\lambda)$ with tridiagonal structure,
 \be
 H^{(2)}(\lambda)=
 \left [\begin {array}{cc} 2&-1-\lambda
 \\{}-1+\lambda&2\end {array}
 \right ]\,,
 \label{dvojka}
 \ee
 \be
 H^{(4)}(\lambda)=
 \left [\begin {array}{cccc} 2&-1&0&0\\{}-1&2&-1-\lambda&0
\\{}0&-1+\lambda&2&-1\\{}0&0&-1&2
\end {array}\right ]\,,
\label{cetyrki}
 \ee
 \be
 H^{(6)}(\lambda)=
 \left [\begin {array}{cccccc}
  2&-1&0&0&0&0
 \\{}-1&2&-1&0&0&0
 \\{}0&-1&2&-1-\lambda&0&0
 \\{}0&0&-1+\lambda&2&-1&0
 \\{}0&0&0&-1&2&-1
 \\{}0&0&0&0&-1&2
\end {array}\right ]\,,\ldots\,.
\label{cetyrkibe}
 \ee
Qualitatively this family of Hamiltonians can be interpreted as a
set of discrete analogues of the exactly solvable ${\cal
PT}-$symmetric square well with a short-range non-Hermiticity
\cite{Quesne}.

The simplicity of our present family of toy Hamiltonians numbered by
their finite matrix dimensions $N=2K=2,4,\ldots$ can be perceived as
the key benefit resulting from our preference of the
non-perturbative Runge-Kutta discretization method. One can
certainly expect that whenever needed, the present $\theta>0$
techniques and constructions will remain applicable also to some
other, less artificial and more phenomenologically oriented
interaction models. Such a transition to more complicated models has
already been shown feasible, in \cite{discrete}, at  $\theta=0$ and
$N = L=\infty$. Similarly, some more-parametric models were shown
tractable by the same method in ref.~\cite{prd}

\subsection{Two-parametric family of  metrics $\Theta^{(N)}(\lambda)$
at  $N=2$ \label{Nje2}}

At $N=2$ and $\lambda=\cos \varphi$  closed formulae are available
not only for the energies $E=E^{(2)}_\pm =2\pm \sin \varphi$ but
also for the norms of the eigenstates $\psi_\pm$. The Hamiltonian
$H^{(2)}(\cos \varphi )$ nicely illustrates the subtle difference
between its right eigenvectors $|\psi_\pm \rangle$ and their
left-eigenstate partners $\bbr \psi_\pm|$ at the same energy (the
latter row vectors are denoted by doubled bras as in~\cite{SIGMA}),
 \be
 |\psi_\pm \rangle \sim \left (
 \ba
 1+\cos \varphi\\
 \mp \sin \varphi
 \ea
 \right )\,,
 \ \ \ \ \
  {\cal T}^{(F)}(\bbr \psi_\pm|):=
 |\psi_\pm \kkt \sim \left (
 \ba
 1-\cos \varphi\\
 \mp \sin \varphi
 \ea
 \right )\,.
 \label{redox}
 \ee
A biorthogonal basis can be formed of these partner eigenvectors.
Thus, the manifestly non-Hermitian matrix $H^{(2)}(\cos \varphi )$
can be reinterpreted as a  matrix which becomes Hermitian in the
{\em ad hoc}, Hamiltonian-dependent Hilbert space of states ${\cal
H}^{(S)}$ endowed with a nontrivial Hermitian-conjugation operation
${\cal T}^{(S)}$ of eq.~(\ref{polaotse}).

The key merit of our $N=2$ example can be seen in the
straightforward availability of {\em all} of its admissible metrics
which vary with two free parameters $t_\pm$ \cite{Ali,SIGMA},
 \be
 \Theta=\Theta^{(2)}(\cos \varphi)=|\psi_+ \kkt\,t_+\,\bbr \psi_+|\,+
 |\psi_- \kkt\,t_-\,\bbr \psi_-|\,.
 \label{jednat}
 \ee
The guarantee of the necessary positivity of these metrics reads
$t_\pm>0$ and holds also, in the similar decoupled form, at all the
higher dimensions $N>2$. After the insertion of  eigenvectors
(\ref{redox}) in (\ref{jednat}) we arrive at our first fully
explicit matrix formula
 \be
 \Theta\, \sim\
 \left (
 \begin{array}{cc}
 (1-\cos \varphi)^2(t_++t_-)&
 (1-\cos \varphi)\sin \varphi(-t_++t_-)\\
 (1-\cos \varphi)\sin \varphi(-t_++t_-)&
 \sin^2 \varphi(t_++t_-)
 \ea
 \right )
 \,.
 \label{opera}
 \ee
Its inspection reveals that up to an irrelevant overall factor it
may be re-written as a strictly equivalent superposition
 \be
 \Theta^{(2)}(\lambda)=\alpha_1\,M_1^{(2)}(\lambda)+\alpha_2\,M_2^{(2)}(\lambda)\,,
 \ \ \ \ \ \ \lambda=\cos \varphi
 \ee
with the two new real free parameters $\alpha_{1}\propto t_++t_-$
and $\alpha_{2}\propto  (-t_++t_-)\sin \varphi$ and with the
following pair of manifestly $\lambda-$dependent sparse-matrix
coefficients,
 \be
 \ \ \ \ \ \ \ \ \ \ \ \
 M_1^{(2)}(\lambda)=
 \left [\begin {array}{cc} 1-\lambda &0\\{}0&1+\lambda\end {array}
\right ]\,, \ \ \ \ M_2^{(2)}(\lambda)=
 \left [\begin {array}{rr} 0&1\\{}1&0\end {array}
 \right ]\,.
 \label{super2}
  \ee
Such a re-parametrization modifies the overall multiplication factor
in $\Theta$ but it still leaves the positivity constraint very
transparent,
 \be
 \alpha_1>0, \ \ \ \ \ \alpha_1^2(1-\lambda^2)>\alpha_2^2\,, \ \ \ \ \
 N=2\,.
 \label{onden}
 \ee
We merely have to choose any $\alpha_2$ from  interval
$(-\alpha_1\sin \varphi,\alpha_1\sin \varphi)$.

The mutual coupling between $\alpha_1$ and $\alpha_2$ is the price
to be paid for the simplification of the $\lambda-$dependence of the
metric. At $N=2$, fortunately, the requirement of a band-matrix form
of $\Theta$ implies that we have to set $\alpha_2=0$ so that the
positivity of the metric will be guaranteed by the elementary
inequality $\alpha_1>0$. At all the higher $N=2K>2$, similarly, the
positivity of the metric will trivially be guaranteed by the set of
requirements $\alpha_1>0$ and $\alpha_2=\alpha_3=\ldots
=\alpha_N=0$. Although the domain of the positivity of the matrix
$\Theta^{(N)}(\lambda)$ will be perceivably larger at $N>2$, its
strict boundary can only be determined numerically in general (cf.
also refs.~\cite{EP} for a very explicit sample study of the
boundaries of the domain of positivity of the metric in the space of
parameters).

\section{Computer-assisted construction of all $N$ metrics  $\Theta^{(N)}(\lambda)$
at $N=4$ and $N=6$ \label{mquattro}}

The mathematical study of the similarity relation
 \be
   \Theta\,H =
   H^\dagger\,\Theta\,
 \label{htot}
 \ee
between a Hamiltonian-type operator $H$ and its adjoint $H^\dagger$
dates back to early sixties \cite{Dieudonne}. In physics, the first
use of such a feature of a sufficiently nontrivial and realistic
Hamiltonian $H\neq H^\dagger$ emerged much later \cite{Geyer}. In
the so called ${\cal PT}-$symmetric quantum mechanics \cite{Carl} an
additional constraint has been accepted by which the metric $\Theta$
is factorized into a product of parity ${\cal P}$ and the so called
charge ${\cal C}$ or quasiparity ${\cal Q}$  \cite{Carl,SIGMAdva}.

In our present paper we shall simply use eq.~(\ref{htot}) as an
(implicit) definition of {\em all the eligible} metrics
$\Theta=\Theta(H)$. Our computer-assisted method of solving this
linear set of algebraic equations for the matrix elements of
$\Theta$ will be straightforward, incorporating also all the
standard requirements imposed upon the metric and listed, say, in
\cite{Geyer}. {\it A priori} we shall not assume the existence of
any other observable like charge or quasiparity. Hence, in our
constructive considerations at finite dimensions only the necessary
Hermiticity $\Theta=\Theta^\dagger$ and positivity $\Theta>0$ of the
metric must and will be required.

\subsection{Ansatz at $N=4$ }

Hamiltonian  $H^{(4)}(\lambda)$ of eq.~(\ref{cetyrki}) offers the
first nontrivial simulation of the non-Hermitian dynamics which is
purely kinetic near its ``distant" boundaries $\pm L$ and which
becomes dynamically nontrivial in the vicinity of the origin. The
coupling $\lambda$ merely connects two points in the middle of the
lattice. The four eigenvalues of matrix $H^{(4)}(\lambda)$ read
 \be
 E_{\pm,\pm}=2\pm \frac{1}{2}\,\sqrt {6-2\,{\lambda}^{2}
 \pm 2\,\sqrt {5-6\,{\lambda}^{2}+{\lambda}^{4}}}\,
 \label{leve}
 \ee
and remain real in the {\em same} interval of couplings $\lambda\in
(-1,1)$ as above. Symbolic manipulations on the computer enable us
to find all the corresponding matrices of the metric
$\Theta^{(4)}(\lambda)$,
 \ben
 \left [\begin {array}{cccc} {\it \alpha_1}\,\left
(1-{\lambda}\right )&{\it \alpha_2}\, \left (1-{\lambda}\right
)&{\it \alpha_3}&{\it \alpha_4}\\\noalign{\medskip}{\it
\alpha_2}\, \left (1-{\lambda}\right )&{\it \alpha_1}\,\left
(1-{\lambda}\right )+{\it \alpha_3}\,\left (1-{\lambda} \right
)&{\it \alpha_4}+{\it \alpha_2}\,\left (1-{{\lambda}}^{2}\right
)&{\it \alpha_3}
\\\noalign{\medskip}{\it \alpha_3}&{\it \alpha_4}+{\it \alpha_2}\,\left (1-{{\lambda}}^{2}
\right )&{\it \alpha_1}\,\left (1+{\lambda}\right )+{\it
\alpha_3}\,\left (1+{\lambda}\right )&{ \it \alpha_2}\,\left
(1+{\lambda}\right )\\\noalign{\medskip}{\it \alpha_4}&{\it
\alpha_3}&{ \it \alpha_2}\,\left (1+{\lambda}\right )&{\it
\alpha_1}\,\left (1+{\lambda}\right )\end {array} \right ]\,.
 \een
They may  be interpreted as the following sum with four variable
real coefficients,
 \be
 \Theta^{(4)}(\lambda)
 =\Theta^{(4)}_{[\alpha_1,\alpha_2,\alpha_3,\alpha_4]}(\lambda)=
 \alpha_1\,M_1+\alpha_2\,M_2+\alpha_3\,M_3+\alpha_4\,M_4\,
 \label{super4}
  \ee
where each component is a sparse matrix carrying a specific
$\lambda-$dependence,
 \ben
 M_1=
 \left [\begin {array}{cccc} 1-\lambda&0&0&0
 \\{}
 0&1-\lambda&0&0\\{} 0&0&1+\lambda&0\\{} 0&0&0&1+\lambda\end {array}\right ]\,,\ \ \ \ \ \
 M_2=
 \left [\begin {array}{cccc}
  0&1-\lambda&0&0\\{}
 1-\lambda&0&1-\lambda^2&0
 \\{}
 0&1-\lambda^2&0&1+\lambda\\{} 0&0&1+\lambda&0\end {array}\right ]\,,
 \een
 \be
 M_3=
 \left [\begin {array}{cccc}0&0&1&0\\{}0&1-\lambda&0&1\\{}
  1&0&1+\lambda&0
 \\{}
 0&1&0&0\end {array}\right ]\,,\ \ \ \ \ \
 M_4=
 \left [\begin {array}{cccc}
 0&0&0&1\\{} 0&0&1&0\\{}
 0&1&0&0\\{}
  1&0&0&0
  \end {array}\right ]\,.
  \label{sihi}
 \ee
The first three items may also be treated as band matrices, i.e., as
a diagonal, tridiagonal and pentadiagonal matrix containing merely
one, two and three nonvanishing diagonals, respectively.

\subsection{Positivity constraint}

As we already indicated in paragraph \ref{Nje2} the diagonal metric
$\Theta^{(4)}_{[1,0,0,0]}(\lambda)\equiv M_1^{(4)}(\lambda)$ remains
safely positive definite inside the open interval of $\lambda\in
(-1,1)$, with  two plus two doubly degenerate eigenvalues
$\mu_{1,2}^{(-)}=1-\lambda$ and $\mu_{1,2}^{(+)}=1+\lambda$. The
remaining three matrices in eq.~(\ref{sihi}) are indefinite. Each of
them possesses a pair of positive and a pair of negative eigenvalues
which are also easily obtainable in closed form.

Once we decide to fix $\alpha_1=1$ and treat the remaining three
parameters $\alpha_2,\alpha_3$ and $\alpha_4$ as small
perturbations, we may also easily establish an allowed range of
these perturbations for which the positivity of the metrics
$\Theta^{(4)}_{[1,\alpha_2,\alpha_3,\alpha_4]}(\lambda)$ remains
robust and guaranteed.

Of course, starting from $N=4$ it is much less easy to describe the
strict position of the $\lambda-$dependent boundary $\partial{\cal
D}$ of the {\em whole} (open) domain ${\cal D}$ of our four real
parameters $\alpha_1\,(=1),\alpha_2,\alpha_3$ and $\alpha_4$ in
which the metric (\ref{super4}) is positive definite. At this
boundary we may expect that the function
$F:=\det\Theta^{(4)}_{[1,\alpha_2,\alpha_3,\alpha_4]}(\lambda)$
(which is equal to the product of the four eigenvalues of the metric
in question) will vanish so that our specification of its zeros it
needed. This task becomes particularly interesting in the maximally
non-Hermitian  dynamical regime, i.e., say, at the couplings
$\lambda = 1 -\varepsilon^2$ where the real variable $\varepsilon$
remains very small and where the (real) energy levels of
eq.~(\ref{leve}) get, pairwise, almost degenerate, with
$E_{0,1}=1\mp \varepsilon / \sqrt{2}+O \left( {\varepsilon}^{2}
\right)$ while $E_{2,3}=3\mp \varepsilon / \sqrt{2}+O \left(
{\varepsilon}^{2} \right)$.

In the zero-order approximation $O \left( {\varepsilon}^{0} \right)$
we reveal, by direct computations, that the determinant $F$ will
vanish whenever  $\alpha_3=\pm \alpha_4$. For illustrative purposes,
let us, therefore, accept this restriction to an exceptional
subspace of parameters and choose the upper sign for the sake of
definiteness. The same argument applied in the next-order
approximation $O \left( {\varepsilon}^{2} \right)$ leads to the
specification of the next quantity $\alpha_2=1 +\alpha_4/2$ and
leaves just the single parameter in the metric unspecified,
$\alpha_4:= \gamma$.

%
\begin{figure}[h]                     
\begin{center}                         
\epsfig{file=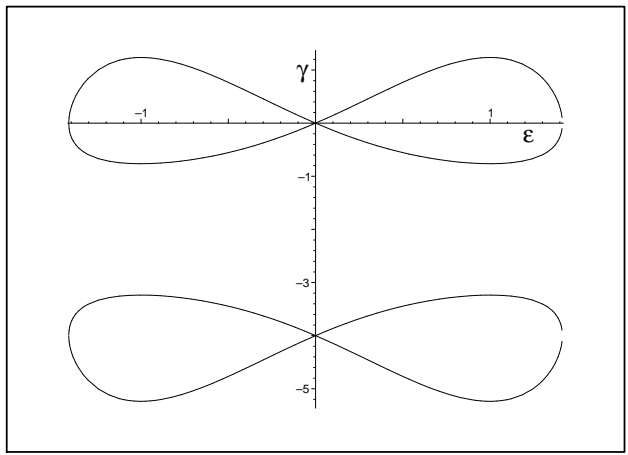,angle=270,width=0.6\textwidth}
\end{center}                         
\vspace{-2mm} \caption{The boundary curve
$\gamma=\gamma(\varepsilon)$.
 \label{firmone}}
\end{figure}

Along the boundary $\partial{\cal D}$ we must have
$F=F(\gamma,\varepsilon)=0$. This is an equation which establishes
an implicit polynomial relationship between $\gamma$ and
$\varepsilon$, i.e., between the ``admissibility boundary"
specifying the span of the positive-definite metrics inside their
selected extremal subset and the strength of the interaction,
respectively. Although the resulting particular curve
$\gamma=\gamma(\varepsilon)$ may be specified by a quadruplet of
explicit formulae for the segments of its boundary,
 \ben
   \gamma=\gamma_\pm^{(\pm)}=
   -2\pm \sqrt {4 \pm 2\,{\varepsilon}\,\sqrt {8-4\,{{\varepsilon}}^{4}
   +{{\varepsilon}}^{6}}+4 \,{{\varepsilon}}^{2}-2\,{{\varepsilon}}^{4}}
   \ ,
 \een
Figure~\ref{firmone} offers a better display of all of its relevant
features. We should emphasize that our considerations are now
non-perturbative so that the independent variable $\varepsilon\in
(-\sqrt{2},\sqrt{2})$ need not stay small. The picture covers its
full range. We may conclude that the interior of all of the four
closed loops in Figure~\ref{firmone} represents the prohibited area
in which the determinant $F$ is negative so that the requirement of
the positivity of the metric matrix is manifestly violated there.

\subsection{Verification of the ansatz at  $N=6$ \label{seisss} }

The $\lambda-$dependence of all of the six eigenvalues of matrix
$H^{(6)}(\lambda)$ may be expressed in closed form as well. They all
prove real (so that the metric $\Theta^{(6)}(\lambda)$ exists) for
all $\lambda \in (-1,1)$. The metric is obtainable either via its
spectral representation \cite{Ali,SIGMA} or, more easily, from
eq.~(\ref{htot}), via its computer-assisted solution. The resulting
matrices $\Theta^{(6)}(\lambda)$ form a six-parametric family
 \ben
 \left [\begin {array}{cccc}
 {\it \alpha_1}\,\left (1-{\lambda}\right )&{\it \alpha_2}
\,\left (1-{\lambda}\right )&{\it \alpha_3}\,\left (1-{\lambda}\right )&\dots \\
{\it \alpha_2}\left (1-{\lambda}\right )&{\it \alpha_1} \left
(1-{\lambda}\right )+{\it \alpha_3}\left (1-{\lambda}\right )&{\it
\alpha_2}\,\left (1 -{\lambda}\right )+{\it \alpha_4}\,\left
(1-{\lambda}\right
)&\ldots\\
{ \it \alpha_3}\left (1-{\lambda}\right )&{\it \alpha_2}\left
(1-{\lambda}\right )+{\it \alpha_4} \left (1-{\lambda}\right
)&{\it \alpha_1}\left(1-{\lambda}\right )+{\it
\alpha_3}\left(1-{\lambda} \right )\left(1-{{\lambda}}^{2}\right
)+{\it \alpha_5}\left (1-{\lambda}\right )& \ldots
\\
{\it \alpha_4}&{\it \alpha_3}\,\left (1-{{\lambda}}^{2}\right )+{
\it \alpha_5}&{\it \alpha_2}\,\left (1-{{\lambda}}^{2}\right
)+{\it \alpha_4}\,\left (1-{{\lambda}}^{ 2}\right )+{\it
\alpha_6}&\ldots
\\
{\it \alpha_5}&{\it \alpha_4}+{\it \alpha_6}&{ \it
\alpha_3}\,\left (1-{{\lambda}}^{2}\right )+{\it \alpha_5}&\ldots
\\
{\it \alpha_6}&{\it \alpha_5}&{\it \alpha_4}&\ldots
\end {array}\right ]=
\een
 \ben
 =\left [\begin {array}{cccc}
 \ldots&{\it
\alpha_4}&{\it \alpha_5} &{\it \alpha_6}\\
\ldots&{\it \alpha_3}\,\left (1-{{\lambda}}^{2} \right )+{\it
\alpha_5}&{\it \alpha_4}+{\it \alpha_6}&{\it
\alpha_5}\\
\ldots&{\it \alpha_2}\,\left (1-{{\lambda}}^{2}\right )+{\it
\alpha_4}\,\left (1-{{\lambda}}^{2}\right )+{ \it \alpha_6}&{\it
\alpha_3}\,\left (1-{{\lambda}}^{2}\right )+{\it \alpha_5}&{\it
\alpha_4}
\\
\ldots&{\it \alpha_1}\left (1+{\lambda}\right )+{\it
\alpha_3}\left (1 +{\lambda}\right )\left (1-{{\lambda}}^{2}\right
)+{\it \alpha_5}\left (1+{\lambda}\right )&{\it \alpha_2}\left
(1+{\lambda}\right )+{\it \alpha_4}\left (1+{\lambda}\right )&{\it
\alpha_3} \left (1+{\lambda}\right
)\\
\ldots&{\it \alpha_2}\left (1+{\lambda} \right )+{\it
\alpha_4}\left (1+{\lambda}\right )&{\it \alpha_1}\left
(1+{\lambda}\right )+{ \it \alpha_3}\left (1+{\lambda}\right
)&{\it \alpha_2}\left (1+{\lambda}\right )
\\
\ldots&{\it \alpha_3}\,\left (1+ {\lambda}\right )&{\it
\alpha_2}\,\left (1+{\lambda}\right )&{\it \alpha_1}\,\left
(1+{\lambda}\right )
\end {array}\right ].
 \een
This formula can be read as a confirmation that in our model an
optimal representation of the general metric will be based on the
use of the general ansatz
 \be
 \Theta^{(N)}(\lambda)=\sum_{j=1}^N\,
 \alpha_j\,M_j^{(N)}(\lambda)\,.
  \label{superdzika}
  \ee
Its $\lambda-$dependence is solely carried by its $N-$dimensional
sparse-matrix coefficients. The matrix elements $(1 \pm \lambda)$
containing the minus-sign always sit in the left upper triangle
(i.e., above the second diagonal) and {\it vice versa}. This
antisymmetry of the sign of ${\lambda}$ with respect to the
reflection of the matrix $\Theta^{(2K)}(\lambda)$ by its second
diagonal simplifies the notation and will hold, incidentally, at all
the integers $K=1,2,\ldots$.

The individual $\lambda-$dependent coefficients $M_j^{(N)}(\lambda)$
are sparse matrices at all the dimensions $N=2K$. For illustration,
the real and symmetric matrix $M_K^{(2K)}(\lambda)$ may be recalled
at $K=4$,
 \ben
 \left [\begin {array}{cccccc} &&&1-{\lambda}&&
 \\
 &&1-{\lambda}&&
 1-{{\lambda}}^{2}&\\
 &1 -{\lambda}&&\left (1-{\lambda}\right )\left
 (1-{{\lambda}}^{2}\right )&&\ldots
 \\
 1-{\lambda}&&\left (1-{\lambda}\right )\left (1-{{\lambda}}^{2} \right )&&\left
 (1-{{\lambda}}^{2}\right )^{2}&\ldots
 \\
 &1-{{\lambda}}^{2}
 &&\left (1-{{\lambda}}^{2}\right )^{2}&& \ldots
 \\
 &&1-{{\lambda}}^{2}&&\left (1+{\lambda}\right )\left (1-{{\lambda}}^{2}\right )& \ldots\\
 &&&
 1-{{\lambda}}^{2}&&\ldots
 \\
 &&&&1+{\lambda}&\end{array}\right ]\,.
 \een
This expression exhibits a seven-diagonal  band-matrix structure.
The similar band-matrix structure will be exhibited by all the
matrices $M_j^{(N)}(\lambda)$ at any $N=2K$ and at all the
subscripts $j$ with the exception of the last one, $j=N$. Precisely
this  property of the matrix components of the metric
(\ref{superdzika}) is responsible for the possibility of the
introduction of a nontrivial fundamental length $\theta$.

\section{Extrapolation towards any $N=2K$ \label{IVaauv}}

Starting from $N = 8$,  matrices $M_j^{(N)}(\lambda)$ get large and
cease to be printable easily. Still, the computer-supported symbolic
manipulations with these matrices remain routine and
straightforward. Moreover, their structure acquires certain features
which enable us to extrapolate their low$-N$ forms to all the even
dimensions $N=2K$ and test the validity of our extrapolations, with
much less effort, afterwards.

One of the most important extrapolation tricks involves the
observation that all of the matrix elements of metric components
$M_k^{(N)}(\lambda)$ form just the collection of the following
sequence of polynomials
 \ben
 P_0=1\,,\ \ \
 P_1^{(\pm)}=1\pm \lambda\,,\ \ \
 P_2=1- \lambda^2\,,\ \ \
 P_3^{(\pm)}=(1\pm \lambda)\,(1- \lambda^2)\,,\ \ \
 \een
 \be
 P_4=(1- \lambda^2)^2\,,\ \ \
 P_5^{(\pm)}=(1\pm \lambda)\,(1- \lambda^2)^2\,,  \ \ \
 P_6=(1- \lambda^2)^3\,,\ \ \  \ldots\,.
 \label{dvacetsest}
 \ee
Their allocation is also not too difficult.

\subsection{Indexing arrays \label{Vab} }

Let us consider the mapping $
 M_k^{(N)}(\lambda) \
 \Longleftrightarrow
 \
S_k^{(N)}$ between our matrix expansion coefficients and certain
arrays of the same size. This specifies, uniquely, each
$\lambda-$dependent matrix coefficient in series~(\ref{superdzika})
using an auxiliary array. At the simplest choice of $N=2$ we have
 \be
 M_1^{(2)}(\lambda)=
 \left [\begin {array}{cc} P_1^{(-)} &0\\{}0&P_1^{(+)} \end {array}
\right ]
 \ \ \ \
 \Longleftrightarrow
 \ \ \ \
 S_1^{(2)}=
 \left [\begin {array}{rr} 1&{\rm }\\{\rm }&1\end {array}
 \right ]\,,
 \label{motykaj}
 \ee
 \be
 M_2^{(2)}(\lambda)=
 \left [\begin {array}{cc} 0&P_0\\P_0&0 \end {array}
\right ]
 \ \ \ \
 \Longleftrightarrow
 \ \ \ \
 S_2^{(2)}=
 \left [\begin {array}{rr} {\rm }&0\\0&{\rm }
 \end {array}
 \right ]\,.
 \label{nemotykaj}
 \ee
At $N=4$ this offers a method of an easy coding or reconstruction of
the four matrices (\ref{sihi}), proceeding via the  following four
indexing arrays $S_j^{(4)}$ at $j=1,2,3,4$, respectively,
 \ben
 \left [\begin {array}{cccc} 1&{}&{}&{}
 \\{}
 {}&1&{}&{}\\{} {}&{}&1&{}\\{} {}&{}&{}&1\end {array}\right ]
 \,,\ \ \ \ \ \
 \left [\begin {array}{cccc}
  {}&1&{}&{}\\{}
 1&{}&2&{}
 \\{}
 {}&2&{}&1\\{} {}&{}&1&{}\end {array}\right ]
 \,,\ \ \ \ \ \
 \left [\begin {array}{cccc}{}&{}&0&{}\\{}{}&1&{}&0\\{}
  0&{}&1&{}
 \\{}
 {}&0&{}&{}\end {array}\right ]
 \,,\ \ \ \ \ \
 \left [\begin {array}{cccc}
 {}&{}&{}&0\\{} {}&{}&0&{}\\{}
 {}&0&{}&{}\\{}
  0&{}&{}&{}
  \end {array}\right ]
 \een
The same observations can be formulated at $N=6$ where the indexing
arrays form the following sextuplet ${S}_j^{(6)}$ at
$j=1,2,\ldots,6$, respectively,
 \ben
 \left [\begin {array}{cccccc} 1&{}&{}&{}&{}&{}
 \\{}
 {}&1&{}&{}&{}&{}\\{} {}&{}&1&{}&{}&{}\\{} {}&{}&{}&1&{}&{}
 \\{} {}&{}&{}&{}&1&{}
 \\{} {}&{}&{}&{}&{}&1
 \end {array}\right ]\,,\ \ \ \ \ \
 \left [\begin {array}{cccccc}{}&1&{}&{}&{}&{}\\{}  1&{}&1&{}&{}&{}
 \\{}
 {}&1&{}&2&{}&{}\\{} {}&{}&2&{}&1&{}\\{} {}&{}&{}&1&{}&1
 \\{} {}&{}&{}&{}&1&{}
 \end {array}\right ]\,,\ \ \ \ \ \
 \left [\begin {array}{cccccc} {}&{}&1&{}&{}&{}\\{} {}&1&{}&2&{}&{}
 \\{} 1&{}&3&{}&2&{}
 \\{} {}&2&{}&3&{}&1
 \\{}
 {}&{}&2&{}&1&{}\\{} {}&{}&{}&1&{}&{}
 \end {array}\right ]\,,\ \ \ \ \ \
 \een
 \ben
 \left [\begin {array}{cccccc} {}&{}&{}&0&{}&{}
 \\{}
 {}&{}&1&{}&0&{}\\{} {}&1&{}&2&{}&0\\{} 0&{}&2&{}&1&{}
 \\{} {}&0&{}&1&{}&{}
 \\{} {}&{}&0&{}&{}&{}
 \end {array}\right ]\,,\ \ \ \ \ \
 \left [\begin {array}{cccccc} {}&{}&{}&{}&0&{}
 \\{}
 {}&{}&{}&0&{}&0\\{} {}&{}&1&{}&0&{}\\{} {}&0&{}&1&{}&{}
 \\{} 0&{}&0&{}&{}&{}
 \\{} {}&0&{}&{}&{}&{}
 \end {array}\right ]\,,\ \ \ \ \ \
 \left [\begin {array}{cccccc} {}&{}&{}&{}&{}&0
 \\{}
 {}&{}&{}&{}&0&{}\\{} {}&{}&{}&0&{}&{}\\{} {}&{}&0&{}&{}&{}
 \\{} {}&0&{}&{}&{}&{}
 \\{} 0&{}&{}&{}&{}&{}
 \end {array}\right ]
 \een
Once we summarize these $N=4$ and $N=6$ computer-generated results
as well as their $N=8$ and $N=10$ descendants we reveal the
existence of the following universal rules.

\begin{itemize}

 \item
For any polynomial $P_n^{(\pm)}$ entering any matrix element $\left
[M_j^{(N)}(\lambda)\right ]_{ik}$ the superscripted $\pm$ sign must
be chosen as $+$ when $i>k$ or as $-$ when $i<k$ or as absent when
$i=k$.

 \item
The indexing symbols  $S_{k}^{(N)}$ as defined by
eqs.~(\ref{motykaj}) and (\ref{nemotykaj}) at $k=1,2$ and $N=2$ get
generalized to any dimension $N=2K$. They always contain either
empty entries or non-negative integer entries ``$n$".

 \item
The numerical value of each entry ``$n$" must coincide with the
value of the subscript $n$ of the related matrix element
$P_n^{(\pm)}$ in the corresponding matrix $M_k^{(N)}(\lambda)$.

\end{itemize}

 \noindent
In the light of these rules the complete determination of the
functions $M_j^{(N)}(\lambda)$ [needed in formula
(\ref{superdzika})] requires just the knowledge of the related
indexing  arrays $S_{j}^{(N)}$. The decoding $S \to M$ using the
above three rules would enable us to reconstruct all the metric
matrices $\Theta^{(2K)}(\lambda)$ via formulae (\ref{superdzika})
and (\ref{dvacetsest}). At any $K$ the indexing arrays
${S}_j^{(2K)}$ with arbitrary subscript $j=1,2,\ldots,2K$ can be
computed in recurrent manner. The description of the details of such
a recipe will be provided in the rest of this section.

\subsection{Recurrences for off-central indexing  arrays $S_j^{(2K)}$, $j \neq K$}

The inspection of the symbols ${S}_j^{(2K)}$ evaluated by the
direct methods at the first few integers $K=1,2,\ldots$ reveals
that at any given $N=2K$ the explicit form of the first $K-1$
matrices ${S}_j^{(2K)}$ with $j=1,2,\ldots,K-1$ may immediately be
deduced from their predecessors ${S}_j^{(2K-2)}$. The core of such
a recurrent construction consists in an enlargement of the
dimension followed by a symmetric attachment of the two $j-$plets
of units $``1"$ in the empty parts of the left upper corner and of
the right lower corner.

The last $K$ matrices ${S}_{2K+1-j}^{(2K)}$ with $j=1,2,\ldots,K$
become formed in similar manner. Their $K$ predecessors
${S}_{2K-1-j}^{(2K-2)}$ must be modified by attaching $j$ zeros
$``0"$ in the right upper corner and in the left lower corner.

In both these ``leftmost-subsequence" and ``rightmost-subsequence"
scenarios,  the results displayed in section~\ref{Vab} offer a
sufficiently instructive illustration of the recipe. They also
indicate that at the ``central" subscript $j=K$ the construction of
the most complicated missing member ${S}_K^{(2K)}$ of the family
must be discussed separately. Although it naturally belongs to the
``leftmost"  subsequence, its $(2K-2)-$dimensional predecessor (to
be denoted as ${\cal L}^{(2K-2)}$) proves {\em different} from the
naively expected matrix ${S}_K^{(2K-2)}$.

\subsection{Recurrences for central indexing  arrays ${S}_K^{(2K)}$ }

The sequence of the ``middle" or ``central" matrices ${S}_K^{(2K)}$
should be treated as exceptional though still generated by a
recurrent recipe. Its idea will rely on the use of specific
predecessor matrices ${\cal L}^{(2K-2)}$. Of course, we shall
proceed in the ``leftmost-subsequence" manner enlarging the
dimension of ${\cal L}$  and filling $K$ units $``1"$ in the left
upper corner and in the right lower corner.

In order  to define the suitable predecessors ${\cal L}^{(2K-2)}$
let us start from the old ``middle" matrix ${S}_{K-1}^{(2K-2)}$
and apply a specific two-step recipe. Firstly we replace each
``old" numerical element in ${S}_{K-1}^{(2K-2)}$ by its successor,
i.e., we replace ``old 0" by ``1", ``old 1" by ``2", etc. In the
second step we form a left-right reflection of the resulting
matrix and arrive at the final form of the necessary predecessor
${\cal L}^{(2K-2)}$ as a result. Thus, at $N=4$ we have the
sequence
 \ben
  {S}_1^{(2)}=
 \left [\begin {array}{cc} 1 &{\rm }\\{}{\rm }&1\end {array}
\right ]\ \to \
 \left [\begin {array}{cc} 2 &{\rm }\\{}{\rm }&2\end {array}
\right ]\ \to \ {\cal L}^{(2)}=
 \left [\begin {array}{cc}  &2\\{}2&{}\end {array}
\right ]\ \to \
  {S}_2^{(4)}\,.
 \een
The recurrent reconstruction of the matrix ${S}_{N/2}^{(N)}$ at
$N=6$  will now result from adding six units ``1" to the auxiliary
predecessor matrix ${\cal L}^{(4)}$ in the formula
 \ben
  {S}_2^{(4)}=
 \left [\begin {array}{cccc}
  {}&1&{}&{}\\{}
 1&{}&2&{}
 \\{}
 {}&2&{}&1\\{} {}&{}&1&{}\end {array}\right ]
  \ \to \
 \left [\begin {array}{cccc}
  {}&2&{}&{}\\{}
 2&{}&3&{}
 \\{}
 {}&3&{}&2\\{} {}&{}&2&{}\end {array}\right ]\ \to \ {\cal L}^{(4)}=
 \left [\begin {array}{cccc}
  {}&{}&2&{}\\{}
 {}&3&{}&2
 \\{}
 2&{}&3&{}\\{} {}&2&{}&{}\end {array}\right ]
 \ \to \
  {S}_3^{(6)}\,
 \een
etc.  We may conclude that the central matrices ${S}_K^{(2K)}$ at
the respective $K=1,2,3,4$ (etc) form the sequence
 \ben
 \left [\begin {array}{cc} 1 &\\&1\end {array}
 \right ], \ \ \ \
 \left [\begin {array}{cccc}
  &1&&\\{}
 1&&2&
 \\{}
 &2&&1\\{} &&1&\end {array}\right ],\ \ \ \
 \left [\begin {array}{cccccc} &&1&&&\\{} &1&&2&&
 \\{} 1&&3&&2&
 \\{} &2&&3&&1
 \\{}
 &&2&&1&\\{} &&&1&&
 \end {array}\right ],\ \ \ \
 \left [\begin {array}{cccccccc} &&&1&&
 &&\\
 &&1&&
 {2}&&&\\
 &1 &&3&&{2}
 &&\\
 1&&3&&4&&{2}
 &\\
 &{2}
 &&4&&3&&1
 \\
 &&{2}&&3&&
 1 &\\
 &&&
 {2}&&1
 &&\\
 &&&&1&&&\end
 {array}\right ]
 \een
(etc). The general pattern of their recurrent production is
obvious and our mathematical construction is complete.

\section{Discussion  \label{uvod}}

Our main mathematical result (\ref{superdzika}) represents the
general metric matrix as its expansion in terms of its sparse-matrix
components $M_{j}^{(N)}(\lambda)$. The latter set has unambiguously
been determined by certain elementary indexing arrays $S_{j}^{(N)}$.
Let us now turn attention to some consequences of such a result in
the context of physics.

\subsection{Fundamental length $\theta$ at large $N$}

Once we pick up the virtually trivial special case where $\lambda
\to 0$ our present toy-model Hamiltonians $H^{(N)}(0)$ become
Hermitian, i.e., tractable as operators of an observable quantity
{\em also} in  ${\cal H}^{(F)}$. Then the elements $|\psi\kt$ of
this friendly Hilbert space  may {\em also} be re-interpreted as
acquiring an entirely standard probabilistic interpretation. This
leads to a rather exotic arrangement (published in \cite{cubic})
where the strict free-motion analogue of our present $\lambda=0$
Hamiltonian has been assigned the family of non-standard metric
operators at $N=\infty$,
 \be
 \Theta^{(Mostafazadeh)}\sim  {\cal I}\cdot
 {\rm cosh} \kappa -  {\cal P}\cdot {\rm sinh} \kappa\,.
 \label{most}
 \ee
Here, symbol ${\cal I}$ represents the identity operator while
${\cal P}$ denotes parity and $\kappa$ is an arbitrary real constant
(cf. eq. Nr. (64) in {\it loc. cit.}). The absence of any
fundamental length has been postulated during the derivation of
eq.~(\ref{most}) in~\cite{cubic}. Thus, in our present Runge-Kutta
discretization scenario the latter result may simply be interpreted,
via identifications $ {\cal I} \to M_1^{(N)}(0)$ and ${\cal P} \to
M_N^{(N)}(0)$, as a particular choice of parameters with vanishing
$\alpha_2=\alpha_3=\ldots=\alpha_{N-1}=0$ and nonvanishing
$\alpha_1\neq 0$ and $\alpha_N\neq 0$. Thus, in our present
terminology one has the vanishing elementary length $\theta=0$ at
$\kappa=0$ and its unbounded alternative $\theta=\infty$ at
$\kappa\neq 0$.

Let us now assume, in contrast, the existence of a finite
fundamental length in our schematic square-well-type example where
the total width $2L$ of the well  need not vary but where the
dimension $N=2K$ itself must always grow to infinity in continuous
limit. In a preparatory step let us assume that $N$ is fixed and
that our toy-model metric $\Theta^{(N)}(\lambda)$ is a
$(2R+1)-$diagonal band matrix. In such a case the natural
elementary-length candidate is $\theta = 2Rh=4LR/(N+1)\sim R$.

In the next step where $N$ starts growing the situation becomes more
complicated because the elementary-length candidate $\theta>0$
should be kept $N-$independent at large $N$.  We arrive at the
conclusion that whenever we try to fix $\theta \neq 0$ and perform
the limiting transition $h \to 0$ (i.e., $N \to \infty$), we must
let the number $\sim 2R+1$ of non-vanishing diagonals {\em grow}
with the dimension. This is the reason why {\em all} the metrics
must be available.

This observation enhances the relevance of our present solvable
model where all the necessary constructions were exact. In some
purely phenomenological applications of the theory with nontrivially
nonlocal metric we may just search for some approximate results and
keep the lattice spacing $h$ fixed. Then we are allowed to fix the
value of $R$ and to select and construct just the particular subset
of metrics with $2R+1$ diagonals. Even in such an entirely pragmatic
setting our present oversimplified example (which admitted the
changes of {\em both} $N$ and $R$) might still prove useful as a
methodical guide.

Alternatively, we might pick up a nontrivial $R>0$ and keep this
integer (i.e., the number of diagonals in $\Theta^{(N)}$) fixed even
during the limiting transition $N \to \infty$ (yielding, formally,
$\theta=0$ of course). Then we still obtain the sequence of metrics
which may converge, typically, to a nontrivial operator
$\Theta^{(\infty)}$ represented by a generalized, momentum-dependent
kernel entering an appropriate operator generalization of our
present, smooth normalization integral~(\ref{laots}), etc.
Naturally, this is a promising but mathematically difficult
possibility which we couldn't have addressed here.

\subsection{The reality of the energies at  large $N$}

From our most elementary toy Hamiltonian $H^{(2)}(\lambda)$ we
deduced the energies most easily, $E=E^{(2)}_\pm =2\pm
\sqrt{1-\lambda^2}$. They remain real in the closed interval of
$\lambda \in (-1,1)$. The  purely numerical analysis of a few
further members $H^{(N)}(\lambda)$ of the family reveals that the
related energy spectra remain real in the same interval of the
couplings $\lambda \in (-1,1)$.

%
\begin{figure}[h]                     
\begin{center}                         
\epsfig{file=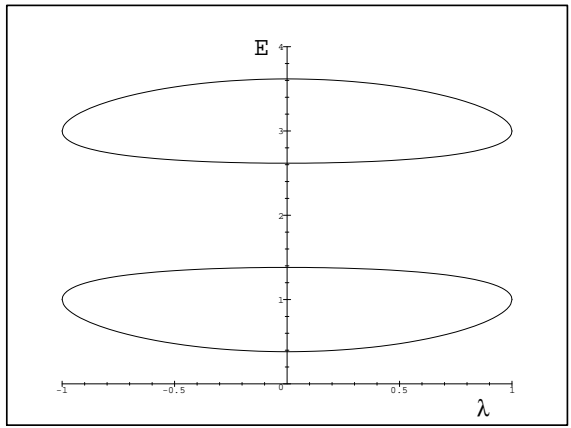,angle=270,width=0.6\textwidth}
\end{center}                         
\vspace{-2mm} \caption{Spectrum of $H^{(4)}(\lambda)$.
 \label{fione}}
\end{figure}

%
\begin{figure}[h]                     
\begin{center}                         
\epsfig{file=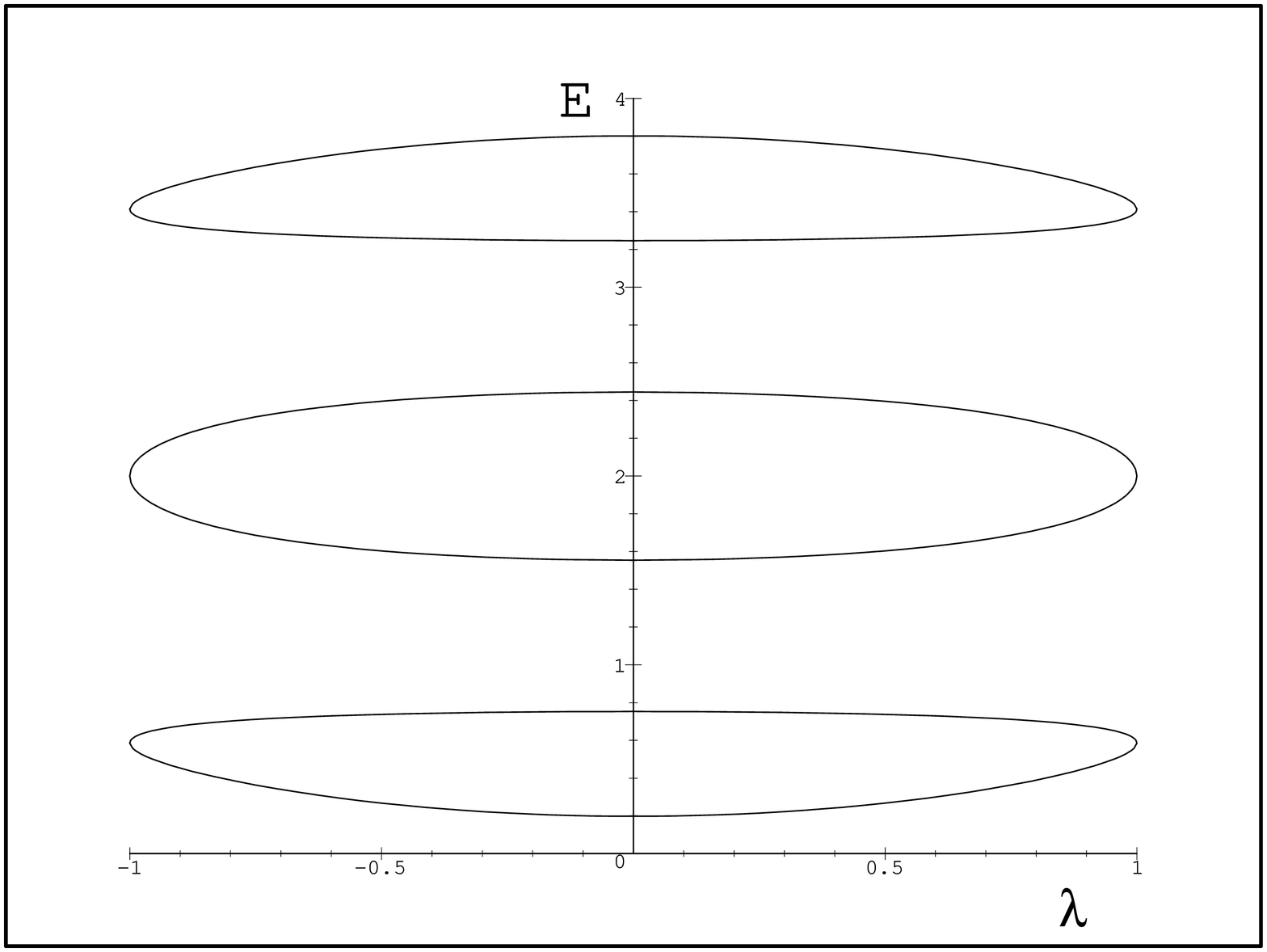,angle=270,width=0.6\textwidth}
\end{center}                         
\vspace{-2mm} \caption{Spectrum of $H^{(6)}(\lambda)$.
 \label{fitwo}}
\end{figure}

Empirically, the validity of this observation is illustrated here by
Figures \ref{fione} and \ref{fitwo}. Sometimes, it may prove useful
to re-parametrize $\lambda =\cos \varphi \in (-1,1)$ with $\varphi
\in (0,\pi)$, therefore (cf., e.g., paragraph \ref{Nje2} above).

One should add that several other features of bound states (e.g.,
the evaluation of matrix elements of some operators of observables)
remain transparent and well illustrated by our discrete short-range
model $H^{(N)}(\lambda)$ and, in particular, by its most elementary
special cases (\ref{dvojka}) -- (\ref{cetyrkibe}).


\subsection{The role of ${\cal PT}-$ and ${\cal CPT}-$symmetry}

On the background of the existence and undeniable physical appeal of
the so called ${\cal PT}-$symmetric differential-equation models as
reviewed by C. Bender \cite{Carl} one reveals that also all the
elements of our sequence of the Runge-Kutta discretized  models can
be incorporated in the same context. Indeed, our Hamiltonian
matrices (\ref{dvojka}), (\ref{cetyrki}), (\ref{cetyrkibe}) etc may
be identified as ${\cal PT}-$symmetric, provided only that we treat
the operator ${\cal T}$ as mediating transposition and that the
parity ${\cal P}$ is represented by the matrix with  units along its
second diagonal,
 \ben
 {\cal P}_{1,N}=
 {\cal P}_{2,N-1}=\ldots=
 {\cal P}_{N,1}=1\,.
 \een
The specific merits of our discrete model involve, furthermore, the
simplicity of mathematical analysis since we avoided the
Rayleigh-Schr\"{o}{dinger-type perturbation expansions reported, in
refs. \cite{Jones,cubic}, as fairly difficult and complicated. In
this context the Runge-Kutta discretization opened a way towards our
straightforward and efficient application of linear-algebraic
techniques.

The post-multiplication or pre-multiplication of any matrix $H$ by
the parity matrix ${\cal P}$ mediates the left-right or up-down
reflection, respectively. We immediately see that our toy
Hamiltonians are not only ${\cal PT}-$symmetric but also, in the
standard terminology of linear algebra, ${\cal P}-$pseudo-Hermitian
\cite{Ali} and ${\Theta}-$quasi-Hermitian \cite{Geyer}. In this
spirit we may recollect ref. \cite{Carl} and try to factorize the
metric, $\Theta= {\cal CP}$. Without the usual additional constraint
${\cal C}^2=I$ this merely defines certain additional,
``protocharge" operators ${\cal C}$. {\it Vice versa}, the
incorporation of the constraint ${\cal C}^2=I$ converts the
protocharge factors into the popular ``charges" \cite{Carl,BBJ}. In
principle, their existence imposes a fairly severe constraint upon
our freedom in the choice of the parameters in the metric
\cite{plb}. At the same time, these constraints may still be
expected to leave some residual freedom in the domain ${\cal D}$ of
admissible parameters $\alpha_j$ \cite{cubic,Kleefeld}.

\section{Summary and outlook \label{summary} }

In the mathematical part of our paper we presented an exact
construction of the most general metric operator
$\Theta^{(N)}(\lambda)$ (i.e., of the most general positive-definite
inner product) for one-parametric non-Hermitian matrix Hamiltonians
$H^{(N)}(\lambda)$ introduced in paper \cite{prd}. The construction
proceeded in two steps. Firstly, the brute-force use of computer
enabled us to generate an exhaustive list of all the admissible
matrices $\Theta^{(N)}(\lambda)$ at the first few (even) matrix
dimensions $N=2K$. In the second step we revealed a clear pattern of
the dependence of these low-dimensional matrices on their dimension
$N$ and on the coupling constant $\lambda$. This enabled us to
arrange each $\Theta^{(N)}(\lambda)$ as a linear superposition of
its $N$ elementary sparse-matrix components $M^{(N)}_j(\lambda)$
with band-matrix structure. In the third step an ansatz for the
latter components has been found and its validity has been confirmed
by extrapolation and its subsequent facilitated verification. In the
final, fourth step the determination of the matrix elements in
$M^{(N)}_j(\lambda)$ was reduced to their indexing via arrays
$S^{(N)}_j$ with integer or empty entries defined via an elementary
recurrent recipe.

Beyond the horizon given by our particular illustrative example we
paid our main attention to the rather serious problem of the
constructive approach to the models with fundamental length $\theta$
and, in particular, to the practical feasibility of the necessary
construction of the related, ``fine-tuned" metric operators
$\Theta$. We tried to overcome the well known difficulties
encountered, in the literature, during perturbation constructions of
the metrics. We found a way how to get rid of at least some of the
current methodical obstructions resulting from the immanent weakness
of perturbation techniques. In this context, the so called
Runge-Kutta approximation techniques were found very productive and
strongly recommendable.

In the parallel physics-motivated discussion of the relevance of our
results in the abstract quantum theory as well as in its various
applications we remind the readers, first of all, that our toy
Hamiltonians $H^{(N)}(\lambda)$ with $N=\infty$ and $\lambda \neq 0$
did already serve as a guide during the recent discussion and
clarification of some conceptual problems concerning the quantum
scattering by non-Hermitian point interactions \cite{prd}. In the
present continuation of their study we succeeded in reconfirming the
relevance of the similar schematic interaction models also in the
theory of bound states.

Our main attention has been paid to conceptual questions again. We
opposed, e.g., the frequently postulated absence of fundamental
length in ${\cal PT}-$symmetric and non-Hermitian models. The
resolution of some concrete technical problems has been found. In
particular we revealed that the extreme simplicity of our model
opens an interesting nonperturbative way towards an innovative and
fully constructive understanding of the emergence of an elementary
length in the quantum system in question. We were also able to add a
few new ideas to the lasting discussions concerning the
interpretation of the presence of a fixed scale $\theta>0$ in
quantum theory. We proposed that in the language using the concept
of metric operators many theoretical considerations may remain
feasible even when a nontrivial constant $\theta$  is introduced.
Last but not least we offered a few arguments supporting the
possibility of using alternatives to the popular and widespread
strategy which connects the ``coordinate-smearing" quantity
$\theta>0$, indirectly and exclusively, to a hypothetical
non-commutativity of classical coordinates.

\vspace{15mm}

\section*{Acknowledgement}

Work supported by the M\v{S}MT ``Doppler Institute" project Nr.
LC06002,  by the Institutional Research Plan AV0Z10480505 and by
the GA\v{C}R grant Nr. 202/07/1307.


\section*{Figure captions}

\subsection*{Figure \ref{firmone}. The boundary curve $\gamma=\gamma(\varepsilon)$.}

\subsection*{Figure \ref{fione}. Spectrum of $H^{(4)}(\lambda)$.}

\subsection*{Figure \ref{fitwo}. Spectrum of $H^{(6)}(\lambda)$.}

\newpage

\section*{Appendix I: Sparse matrices $M^{(N)}(\lambda)$ at  $\lambda=0$ }

The formal structure of the complete sets of metrics
$\Theta^{(N)}(\lambda)$ collected at the smallest dimensions $N\leq
6$ gets fully transparent in the vanishing-potential limit
$\lambda\to 0$ when all the Hamiltonians become Hermitian in the
$N-$dimensional Hilbert spaces ${\cal H}^{(F)}$. The textbook choice
of the metric looks unique because people tacitly assume that there
exists no nontrivial fundamental length in the theory (cf., e.g.,
ref. \cite{cubic}). In our present notation such an option coincides
with the special case where $\theta=0$, $\alpha_1>0$ (i.e., say,
$\alpha_1=1$) and $\alpha_2=\alpha_3=\ldots =\alpha_N=0$.

Whenever we intend to build the theory where the choice of
$\theta>0$ sets a nontrivial length scale, Hamiltonian $H$ need not
be non-Hermitian in ${\cal H}^{(F)}$. In our toy model at
$\lambda=0$, in particular, we may still define a non-Dirac metric
using formula (\ref{superdzika}) with some nonvanishing values of
parameters $\alpha_2$ and $\alpha_3$ etc. It is only necessary to
guarantee that the metric matrix itself remains positive (for
illustration, recollect inequality (\ref{onden}) which specifies the
full allowed range of parameters at $N=2$).

Although our models $H^{(N)}(\lambda)$  admit a nontrivial
fundamental length $\theta>0$ even in their square-well limit
$\lambda=0$ of paper \cite{sqw}, possible physics represented by
such an extreme example looks rather artificial, Still, its
methodical merits are remarkable. Firstly, the coefficient matrices
$M=M_j^{(N)}(0)$ become solely filled by the matrix elements 0 or 1.
Secondly, their knowledge may prove useful for coding the indexing
arrays in computer-assisted manipulations. Thirdly, the simplicity
of the model implies that the $j-$th member of the sequence
$M_j^{(N)}(0)$ can be defined by the following closed formula,
 \be
 \left (M_j^{(N)}\right )_{ik}(0)= 1
 \ \ \ {\rm iff}\ \ \ i-k=m\,,\ \ \ N+1-i-k=n\,,\ \ \
 \ \ \ \ \ \ \
 \label{anza}
 \ee
 \ben
 \ \ \ \ \
 m=j-1, j-3, \ldots, 1-j\,,
 \ \ \ \ \ \ \ \
 n=N-j, N-j-2, \ldots, j-N\,
 \een
and the verification of validity of this formula by its direct
insertion in eq.~(\ref{htot}) is very quick. Fourthly, the existence
of this and related formulae may prove useful for perturbation
constructions in weakly non-Hermitian dynamical regime where
$|\lambda| \ll 1$.

\section*{Appendix II: Our toy Hamiltonians at large $N$}

At a sufficiently large $N$ our particular one-parametric
Hamiltonians $H^{(N)}(\lambda)$ may be interpreted as  discrete
versions of a differential operator with a point interaction 
localized in the origin. For a deeper understanding of such a
correspondence let us  abbreviate $2-h^2E=\cos \epsilon$ as usual
\cite{sqw}. We may then treat $\epsilon\in (0,\pi)$ as a new energy
variable and visualize the wave functions $\psi(x)$ with $x \neq 0$
as satisfying a free-motion equation complemented by the respective
left and right initial conditions $\psi(-L)=\psi(L)=0$. At a fixed
$L$ and in the $N=2K\gg 1$ approximation these free-motion-like
solutions must be further restricted by the pair of
$\lambda-$dependent constraints near the origin,
 \be
 (1+\lambda)\,{\psi(x_{1})-2\cos
 \epsilon\,\psi(x_{{0}})+\psi(x_{-1})}=0\,,
 \label{diskra}
 \ee
 \be
 {\psi(x_{{2}})-2\cos
 \epsilon\,\psi(x_{{1}})+(1-\lambda)\, \psi(x_{{0}})}=0\,.
 \label{diskrb}
 \ee
In the limit $h \to 0$ we may expect the emergence of a
discontinuity in $\psi(x)$ at $x=0$. Even at all the finite $N \sim
1/h \gg 1$ the wave functions remain well represented by their
respective one-sided Taylor series near $x=0$ so that
eqs.~(\ref{diskra}) and (\ref{diskrb}) may be interpreted as a
matching condition. Once we  return to the original energy variable
$h^2E=2-2\cos \epsilon\equiv F$ and insert the truncated expansions
 \ben
 \psi(x_{-1})=\psi_L(0) -\frac{3}{2}\,h\,\psi_L'(0)
  +{\cal O}(h^2)\,,\ \ \ \ \
 \psi(x_{{0}})=\psi_L(0) -\frac{1}{2}\,h\,\psi_L'(0)
  +{\cal O}(h^2)\,,
 \een
 \ben
 \psi(x_{1})=\psi_R(0) +\frac{1}{2}\,h\,\psi_R'(0)
  +{\cal O}(h^2)\,,\ \ \ \ \
 \psi(x_{2})=\psi_R(0) +\frac{3}{2}\,h\,\psi_R'(0)
  +{\cal O}(h^2)\,
 \een
in eqs.~(\ref{diskra}) and (\ref{diskrb}), a straightforward algebra
leads to the following elementary condition
 \be
 \frac{h}{2}\,
 \left (
 \begin{array}{cc}
 -(1+\lambda)&F+1\\
 -(F+1)&1-\lambda
 \ea
 \right )\,
 \left (
 \ba
 \psi_R'(0)\\
 \psi_L'(0)
 \ea
 \right )=
 \left (
 \begin{array}{cc}
 1+\lambda&F-1\\
 F-1&1-\lambda
 \ea
 \right )\,
 \left (
 \ba
 \psi_R(0)\\
 \psi_L(0)
 \ea
 \right )\,
 \label{mbc}
 \ee
which matches the wave functions and their derivatives in the
origin.

In the domain of sufficiently small $h>0$ the latter relation is
equivalent to the original constraints (\ref{diskra}) and
(\ref{diskrb}). We may conclude that at all the  nonvanishing small
$h>0$ our conditions (\ref{mbc}) leave our interaction in the origin
translucent and manifestly energy-dependent.

Various special cases of our $N\gg 1$ bound-state model may be
studied noticing, for example, that the energy-dependence disappears
in the low-excitation regime where the quantity $F=h^2E$ remains
negligible. At a generic energy $F>0$ the above set of $h>0$
solutions must be complemented by the two additional, anomalous
bound states emerging at the two exceptional energies $F =F_\pm = 1
\pm \sqrt{1-\lambda^2}$ which make the coefficient matrix singular.
At these energies, both the values of $\psi_{R,L}(0)$ are, in
general, non-vanishing and firmly determined by our choice of the
two derivatives $\psi'_{L}(-L)$ and $\psi'_{R}(L)$ at $x=\mp L$,
respectively. Thus, both the left and right branches of our two
exceptional bound states are obtained by the same matching in the
origin as above. Their specific feature is that at both our
exceptional energies $F_\pm$ the two lines of eq.~(\ref{mbc})
degenerate, at all the sufficiently small $h \approx 0$, to the
single constraint $\psi_R(0)\,\sqrt{1+\lambda}\pm
\psi_L(0)\,\sqrt{1-\lambda}=0$. Thus, manifestly asymmetric wave
functions are obtained.

In the continuum limit $N\to \infty$ our sequence of the matrix
Hamiltonians $H^{(N)}(\lambda)$ can be reinterpreted as a series of
dynamical models which converge to a  specific differential equation
with a point interaction potential in the origin. It is easily seen
that at a generic energy $E$ the $h\to 0$ limit of eq.~(\ref{mbc})
leads to the vanishing $F ={\cal O}(h^2)$ so that the
above-mentioned ``exceptional" solutions disappear from the
spectrum. At any $\lambda \neq 0$, only the elementary opaque-wall
constraint $\psi_R(0)=\psi_L(0)=0$ survives and leads, say, to the
two independent series of bound-state solutions which live solely on
the left or right half-interval of $x$, respectively.

\end{document}